\documentclass[reprint,showpacs,
amsmath,amssymb,aps,prb,floatfix,numerical]{revtex4-1}
\usepackage{natbib}
\usepackage{graphicx}% Include figure files
\usepackage{booktabs}
\usepackage{multirow}
\usepackage[table,xcdraw]{xcolor}
\usepackage{dcolumn}% Align table columns on decimal point
\usepackage{bm}% bold math
\usepackage{siunitx}
\usepackage[version=3]{mhchem}
\usepackage[british]{babel}
\usepackage{hyperref}% add hypertext capabilities
\usepackage{amsmath}
\usepackage{epstopdf}

\newcommand{\abs}[1]{\left|#1\right|}

\newcolumntype{C}{>{\centering\arraybackslash}X}
\begin{document}
    \preprint{APS/123-QED}
    \title{Polarization branches and optimization calculation strategy applied to $ABO_3$ ferroelectrics}
    \author{Lucian D. Filip}
    \email[Corresponding author: ]{lucian.filip@infim.ro}
    \author{Neculai Plugaru}
    \affiliation{National Institute of Materials Physics, Atomistilor str., nr. 405A, P.O. Box MG7, Magurele, Bucharest, Romania}
\begin{abstract}
Berry phase polarization calculations have been investigated for several ferroelectric materials from the point of view of practical calculations. 
It was shown that interpretation of the results is particular to each case due to the multivalued aspect of polarization in the modern theory. 
Almost all of the studied examples show ambiguous polarization results which can be difficult to solve especially for super-cells containing large number of atoms. 
For this reason, a procedure has been proposed to minimize the number of calculations required to produce an unambiguous polarization result from Berry phase polarization investigations. 
\end{abstract}
\maketitle
%--------------------------------------------------------------------------------------------------------
\section{\label{sec:Intro}Introduction}
Ferroelectric materials, have been under intense research for the better part of the last three decades due to the tremendous integration potential into applications ranging from high-density non-volatile memories to solar cells and logic gates \cite{scott1989,hidaka1992,glinchuk2009,kumari2015,rong2016,scott2007,choi2009,garcia2009,huang2010,yuan2011,liu2014,chen2015,kim2016}. 
Most ferroelectrics belong to the \emph{perovskite} $ABO_3$ family which is a versatile group of materials that can be obtained through a wide range of synthesis methods \cite{moure2015}. 
One important property for real life applications is the large polarization value obtained in some ferroelectric compounds such as $Pb(Zr_{x}Ti_{1-x})O_3$ (PZT) thin films (around \SI{1}{\coulomb\per\metre\squared}) \cite{pintilie2007}. 
Unfortunately these high polarization values are mostly obtained in lead-containing materials which have a rather large environmental footprint. 
For this reason, there is a sustained experimental and theoretical effort to find lead-free perovskites that ideally retain the high polarization value but have a much lower toxicity. 
The theoretical approach for this search is based on high throughput calculations using automated scripts searching for particular material properties (such as a high polarization value) \cite{kundu2015,volonakis2016,filip2016,filip2016a}. 
The most reliable theoretical model to date for computing the spontaneous polarization is an implementation of the Berry phase (BP) formalism \cite{berry1984} adapted by Resta \cite{resta1992} in 1992 to correctly define and calculate the bulk polarization value of a ferroelectric. 
It was later implemented by King-Smith and Vanderbilt \cite{kingsmith1993} in first-principles density-functional theory numerical calculations routine and together have become what is now called the \emph{modern theory of polarization}. 
This method has been used to compute polarization values for various bulk ferroelectric materials. 
However, special attention is required for each studied material. 
The difficulty lies behind the definition of the spontaneous polarization within the modern theory of polarization as the time integral of the electric current appearing when the system is distorted adiabatically from a reference centrosymmetric state (CS) towards a final ferroelectric state (FS) \cite{resta1992,resta1993a,resta1993b,resta1994,rabe2007}. 
Therefore, in order to calculate the spontaneous polarization, one needs the polarization values for at least two different system states (CS and FS). 
Henceforth the spontaneous polarization is the difference between the two. 
Therefore such an approach is fundamentally different than previously used models where the bulk polarization was viewed as a collection of neatly arranged electrical dipoles \cite{rabe2007}. 
In addition, the calculated polarization is \textbf{\underline{not}} single valued but a multivalued function of the system state; it can only be obtained up to an integer number (indexing polarization branches) multiplied by a certain geometrical system constant \cite{resta1992,resta1993a,resta1994,rabe2007}. 
This intricacy complicates the interpretation of numerical results considerably, since the obtained polarization values for the required CS and FS may not have the same indeterminacy (\emph{i.e.}\ they may belong to different polarization branches). 
As such the difference in polarization of the CS and FS is multivalued contradicting experimental spontaneous polarization measurements. 
This ambiguity is corrected by repeating calculations for a large number of intermediary system states, so as to identify the polarization branches of the CS and FS. 
However, the procedure is system dependent and can also lead to incorrect results if the proper corrections are not used, as pointed out by Neaton \emph{et. al} \cite{neaton2005}. 

It is at this point where a distinction between the analytical theoretical description of the modern theory of polarization and its numerical implementation and usage, should be made. 
While the BP theory of polarization is general to all materials, the interpretation of results obtained from the numerical implementations are strongly system dependent. 
This fact implies that almost always one must resort to calculating the polarization for multiple intermediary system states which can be very time consuming especially for larger systems. 

The present paper aims to illustrate the BP polarization theory from the perspective of practical density functional theory \cite{hohenberg1964} calculations and to introduce a strategy for choosing the optimum calculation points required to resolve the polarization ambiguity in a minimum number of steps. 
The manuscript is structured as follows: in the next section, the materials used in this study are presented, followed by the computational details in Section \ref{sec:compdet}. 
Berry phase calculations are briefly described in Section \ref{sec:bpusage} and the polarization calculations results for the materials introduced in Section \ref{sec:mat} are individually discussed in Section \ref{sec:calc}. 
The proposed optimization strategy is introduced in Section \ref{sec:optimization} using the example of a larger system. 
The manuscript will be concluded with a discussion of all the results obtained in this study and the implemented strategy in Section \ref{sec:res}. 
%--------------------------------------------------------------------------------------------------------
\section{\label{sec:mat}Materials and structures}
\begin{figure}[h]
    \centering
    \includegraphics[width=\columnwidth]{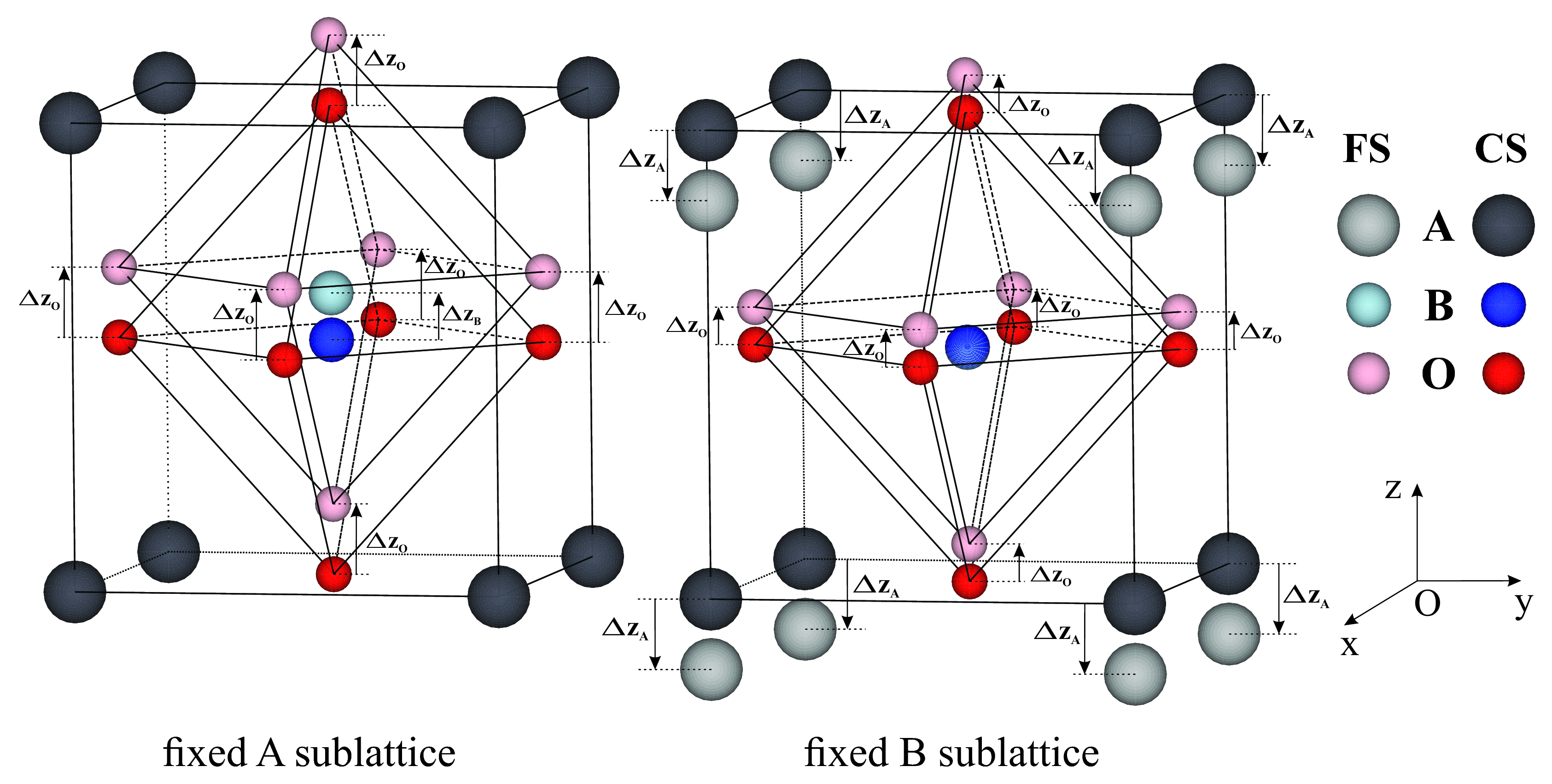}
    \caption{Schematic of the tetragonal unit-ell a prototypal $ABO_3$ ferroelectric perovskite. Two possible distortion paths are exemplified with respect to the A and B sublattices, respectively.}
    \label{fig:diag}
\end{figure}

The application of the BP polarization theory can be better understood if practical calculations are discussed. 
For this reason, in this study, four materials have been used to exemplify various aspects of polarization calculations: $BaTiO_3$ (BTO), $PbTiO_3$ (PTO), $KNbO_3$ (KNO) and $Pb(Zr_{0.25}Ti_{0.75})O_3$ (PZT). 
The choice for the first three materials has been motivated by the simplicity of their respective unit-cells which allowed for a large number of calculations to be performed for intermediary system distortion states. 
The symmetry of the systems is tetragonal and they each have five atoms in the unit-cell (see Figure \ref{fig:diag}). 
The lattice parameters and atomic coordinates for the first three materials have been summarized in Table \ref{tab:1}. 

\begin{table}
    %    \renewcommand{\arraystretch}{1.3}
    %    \centering
    \begin{ruledtabular}
        \begin{tabular}{@{}ccccccc@{}}
            \multicolumn{2}{c}{} a (\AA) & c (\AA) & $z_{A}$ & $z_{B}$ & $z_{O_{1}}$ & $z_{O_{2,3}}$ \\
            $BaTiO_3$ & 3.9925 & 4.0365 & 0.00 & 0.4785 & 0.0253 & 0.5105 \\
            $PbTiO_3$ & 3.8775 & 4.2070 & 0.00 & 0.5380 & 0.1166 & 0.6211 \\
            $KNbO_3$  & 3.9970 & 4.0630 & 0.00 & 0.4770 & 0.0170 & 0.519  \\
        \end{tabular}
        \caption{Lattice parameters and atomic coordinates of the optimized $PbTiO_3$, $BaTiO_3$ and $KNbO_3$ unit-cells.}
        \label{tab:1}
    \end{ruledtabular}
\end{table}
$Pb(Zr_{0.25}Ti_{0.75})O_3$ has been used to exemplify BP polarization calculations for a larger system. 
The 40 atom super-cell was obtained by arranging 8 PTO unit-cells in a $2\times2\times2$ grid and replacing two $Ti$ atoms with $Zr$. 
Being a larger system the polarization calculations took considerably longer time to complete compared to calculations for the smaller studied systems. 
While the relative time it takes for a polarization calculation to finish is not relevant when individual materials are tested, it becomes important when multiple such calculations are performed for an entire range of different materials. 
%--------------------------------------------------------------------------------------------------------
\section{\label{sec:compdet}Computational details}
All calculation for the present study were performed using the generalised gradient approximation (GGA) density functional theory \cite{hohenberg1964} as implemented in the Quantum Espresso suite \cite{giannozzi2009}. 
Projected augmented-wave (PAW) Perdew-Burke-Ernzerhof pseudo-potentials optimized for solids (PBEsol) \cite{perdew2008}, from the THEOS library \cite{theos}, have been used for all the numerical results obtained in this study. 
However, it should be noted that while small differences can be obtained for the value of the spontaneous polarization, the ambiguities generated by the multivalued aspect remain. 
For this reason the precision of the numerical results in this paper is less important since they were used for visualisation of the peculiarities involved with the Berry phase polarization calculations. 

For the first three materials, a kinetic energy cut-off of $\SI{100}{Ry}$ was chosen and the Brillouin zone was sampled in a $5\times5\times5$ Monkhorst-Pack \cite{monkhorst1976} uniform k-point grid for all self-consistent calculations. 
The polarization calculations (non-self-consistent) were performed with a $5\times5\times20$ k-point grid. 
For the BTO and PTO materials, structural optimization was performed starting from experimental data found in the literature \cite{wang2010,saghiszabo1998}. 
The optimized lattice parameters were obtained by minimizing the total energy with respect to the volume of the unit cell and the final atomic coordinates were obtained by relaxing the internal coordinates until the Hellmann-Feynman forces are converged to \SI{2.6e-9}{\eV\per\angstrom}. 
The KNO case has been treated using the same lattice parameters and atomic coordinates as obtained by Dall'Olio \emph{et. al} \cite{dallolio1997}. 

The $Pb(Zr_{0.25}Ti_{0.75})O_3$ material was also investigated as an example of a larger system described by a super-cell formed with 8 PTO unit-cells disposed in a $2\times2\times2$ grid where 2 $Ti$ atoms have been replaced by $Zr$. 
The lattice parameters of the PZT super-cell are twice the ones obtained for PTO (see Table \ref{tab:1}). 
The atomic coordinates were further relaxed to account for the presence of the two $Zr$ atoms. 
Since the larger system will take considerably longer time for each polarization calculation run to complete, it was used to highlight the consequences of randomly choosing intermediary system distortion states between CS and FS in order to solve the polarization ambiguity. 
Lastly, as a language convention, a polarization calculation run is a set of two consecutive calculations for a given system state: first a self-consistent calculation on a uniform k-point grid, followed by a non-self-consistent calculation with an increased number of k points along the axis where the polarization is calculated. 
%--------------------------------------------------------------------------------------------------------
\section{\label{sec:bpusage}Berry phase primer and usage}
The modern theory of polarization defines ferroelectric polarization as the time integral of the current that appears through the sample when the studied system is adiabatically distorted from a reference state to a final ferroelectric state. 
This is valid as long as the system remains insulating in all the intermediate states along the distortion path \cite{rabe2007}. 
A visual representation of two such processes are given in Figure \ref{fig:diag} where the tetragonal $ABO_3$ system is distorted from the CS to the FS along the z axis with respect to the A and B sublattices, respectively. 
For all materials studied in this paper, the polarization direction is parallel to the z axis of the unit-cell and thus the systems will only be distorted along this axis. 
For the cases shown in Figure \ref{fig:diag} we can write the system distortion as a linear function of a dimensionless parameter $\lambda$:
\begin{equation}
z_{i}(\lambda)=z^{CS}_i+(z^{FS}_i-z^{CS}_i) \, \lambda,
\label{eq:distort}
\end{equation}
where the index $i$ spans all the atoms in the unit cell and $\lambda \in [0,1]$. 
The coordinates $z^{CS}_i$ and $z^{FS}_i$ represent the $z$ coordinates of atom $i$ for the centrosymmetric and the ferroelectric state, respectively, while $z_{i}(\lambda)$ is the coordinate of the same atom $i$ for an intermediary distortion $\lambda$. 
This means that all the atoms in the unit-cell will be moved together by a fraction $\lambda$ of their corresponding final displacements $\Delta z_i=z^{FS}_i-z^{CS}_i$. 
Following the definition of polarization in the modern theory, the difference between the values obtained for the FS and CS should represent the spontaneous polarization of the studied system. 
\begin{equation}
P_s=P_{FS}-P_{CS},
\label{eq:pss}
\end{equation}
where, $P_s$ is the measurable spontaneous polarization, $P_{FS}$ the computed polarization in the ferroelectric state and $P_{CS}$ the computed polarization in the reference state (in our case, the centrosymmetric state). 
This is a measurable quantity and the modern theory mimics the generic experimental measurement method, where the system state is switched between the two stable ferroelectric states using an external electric field along a hysteresis cycle. 

However, since the polarization is a multivalued function, Eq. \eqref{eq:pss} does not strictly represent the spontaneous polarization \cite{rabe2007,resta1993a}. 
Indeed it can be shown that polarization is only well-defined modulo a polarization quantum given by: $P_q=\frac{e \, \bm{R}}{\Omega}$, where $\Omega$ is the unit-cell volume, $e$ is the unit charge and $\bf{R}$ is any lattice vector. 
This means that the calculated polarization for any system state $P_{state}$ is actually an entire family of values separated by integer multiples of $P_q$ (also called branches) and given by \cite{resta1993a,rabe2007}:
\begin{equation}
\bm{P}_{state}=\bm{P}_{Berry}+n\,\bm{P_q}, \, n \in \mathbb{Z},
\label{eq:pstate}
\end{equation}
where, $P_{Berry}$ is the calculated polarization and $n$ is an integer indexing the polarization branch \cite{resta1993a,rabe2007}. 
It is now clear that, if the polarization values for the CS and FS do not belong to the same branch, then the spontaneous polarization will continue to have the form given in Eq. \eqref{eq:pstate}, contradicting experimental findings. 

\begin{figure}[h]
    \centering
    \includegraphics[width=\columnwidth]{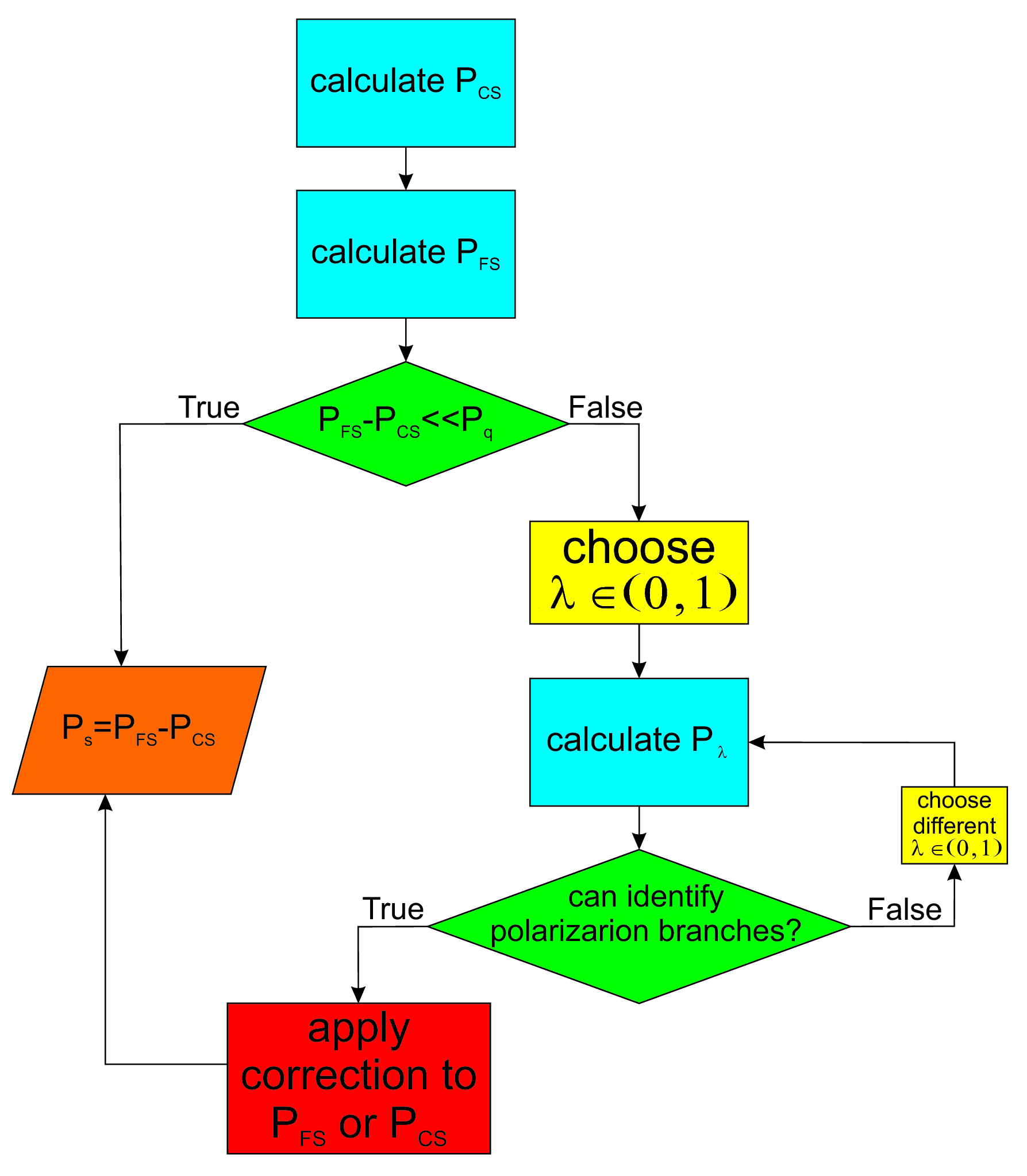}
    \caption{Berry phase polarization method calculation work-flow. The blue rectangles represent polarization calculation runs, green diamonds are decisional steps. Red, yellow and orange rectangles are regular arithmetic operations.}
    \label{fig:berrywfl}
\end{figure}

Eq. \eqref{eq:pstate} is not only an important consequence of the modern approach to ferroelectric polarization but also the key to the correct application of the numerical implementation of this theory for practical cases. 
From the theoretical point of view, one can identify if the polarization values for the CS and FS belong to the same branch by looking at their difference and comparing it to the polarization quantum \cite{resta1994,kingsmith1993,rabe2007}. 
If the difference is much smaller than the polarization quantum then no ambiguity has appeared and Eq. \eqref{eq:pss} is valid. 
For the case when the difference is comparable to the polarization quantum, intermediary system distortions should be considered in order to clarify what branch the two polarization values belong to and which corrections should be made. 
However, there is no clear indication what it means for the polarization difference to be ``much smaller" and it will be shown that this comparison cannot be used effectively to solve the ambiguity. 
Figure \ref{fig:berrywfl} summarizes the calculation steps in order to obtain the spontaneous polarization using the Berry phase polarization method. 
%--------------------------------------------------------------------------------------------------------
\section{\label{sec:calc}Numerical calculations and polarization branches}
%--------------------------------------------------------------------------------------------------------
\subsection*{$BaTiO_3$}
\begin{figure}[h]
    \centering
    \includegraphics[width=\columnwidth]{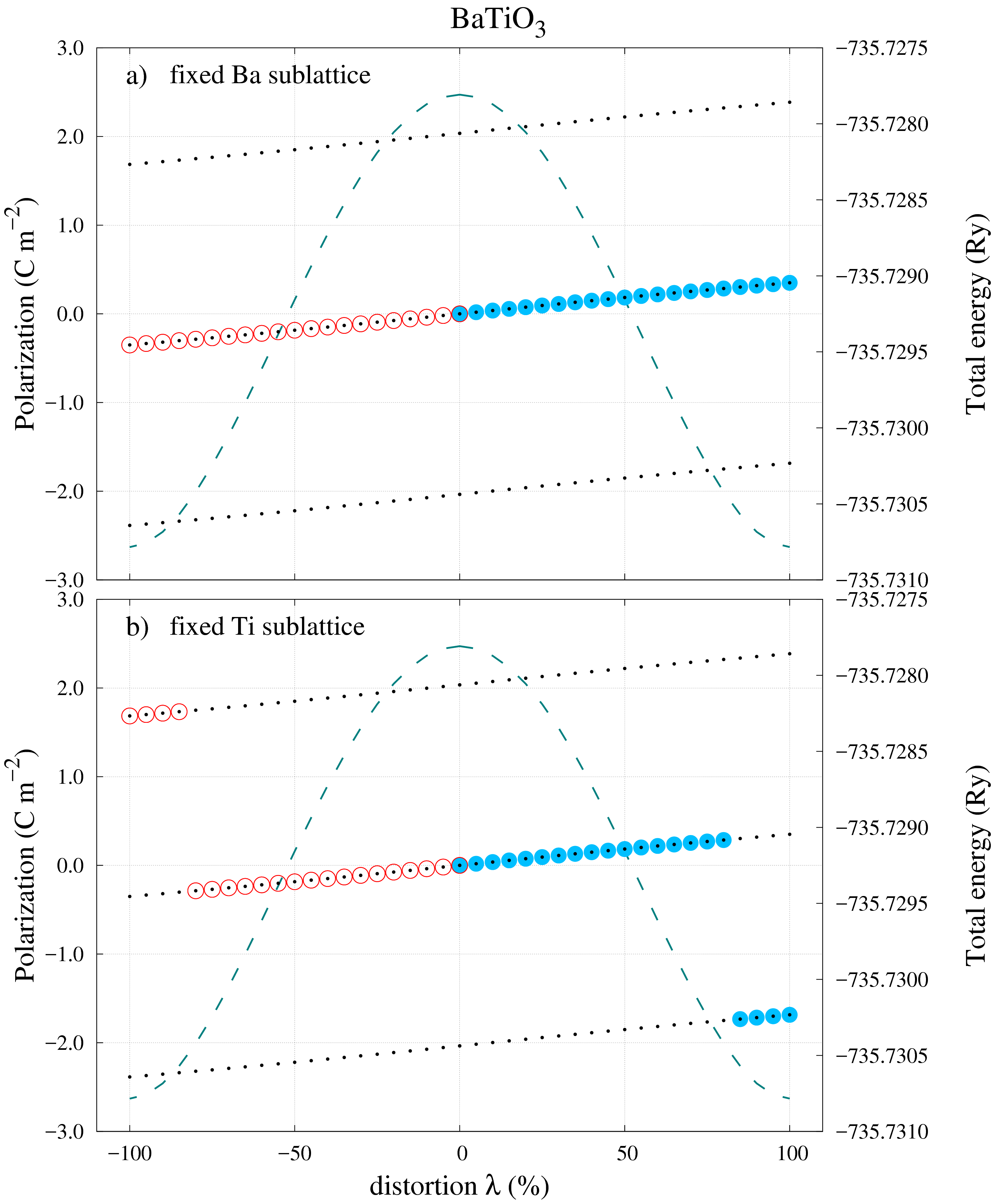}
    \caption{Polarization lattice for $BaTiO_3$ (black dots given by Eq. \eqref{eq:pstate}) with distortion from the CS to the ``DOWN" FS (red empty circles) and from the CS to the ``UP" FS (solid blue circles) with respect to the $Ba$ sublattice a) and with respect to the $Ti$ sublattice b). The distortion parameter $\lambda$ was modified in $5\%$ steps, with the negative values for the ``DOWN" intended for a better clarity of the figure. The green dashed line is the total energy as a function of distortion showing a continuous variation for each distortion path.}
    \label{fig:bto}
\end{figure}

The first choice of distortion path shown in Figure \ref{fig:bto}a) returns the ideal result: All the calculated polarization values belong to a single polarization branch. 
It is then clear that by applying Eq. \eqref{eq:pss} with the FS and CS values one obtains the correct polarization value of $\SI{0.351}{\coulomb\per\metre\squared}$. 
Using the checkpoints at the end of Section \ref{sec:bpusage}, the values for the ``UP" directions are as follows: $P_{CS}=\SI{0}{\coulomb\per\metre\squared}$, $P_{FS}=\SI{0.351}{\coulomb\per\metre\squared}$ and $P_{q}=\SI{2.036}{\coulomb\per\metre\squared}$. 
The difference between the FS and CS values is almost 6 times smaller than the polarization quantum and this could have been considered sufficient to conclude that an unambiguous result can be obtained without any other intermediary system distortions. 
However, the situation changes drastically if the unit-cell of the system is now distorted with respect to the B atom site (see Figure \ref{fig:bto}b)). 
From the CS to around $80\%$ distortion, the calculated polarizations are neatly arranged on the same branch just like in the previous example. 
Nevertheless for the rest of the $20\%$ left to the final FS the values jump suddenly on a different branch. 
Analysing the end values of interest, one obtains: $P_{CS}=\SI{0}{\coulomb\per\metre\squared}$ and $P_{FS}=\SI{-1.685}{\coulomb\per\metre\squared}$, while the polarization quantum remains the same, $P_{q}=\SI{2.036}{\coulomb\per\metre\squared}$. 

According to the calculations steps outlined in Figure \ref{fig:berrywfl} the difference is now $\SI{-1.685}{\coulomb\per\metre\squared}$ which is almost the same as $P_q$ in absolute value. 
This means that an ambiguity has arisen and a correction is needed. 
In this case, with that many calculated values, it is easy not only to identify where the calculated polarization values reside on their corresponding branch but also what correction should be used. 
For this case, by adding one polarization quanta to $P_{FS}$ will bring its value on the same branch as $P_{CS}$ (for the ``UP" direction). 
The spontaneous polarization in Eq. \eqref{eq:pss} becomes: $P_s=P_{FS}+P_q-P_{CS}=\SI{0.351}{\coulomb\per\metre\squared}$. 
%--------------------------------------------------------------------------------------------------------
\subsection*{$PbTiO_3$}
\begin{figure}[h]
    \centering
    \includegraphics[width=\columnwidth]{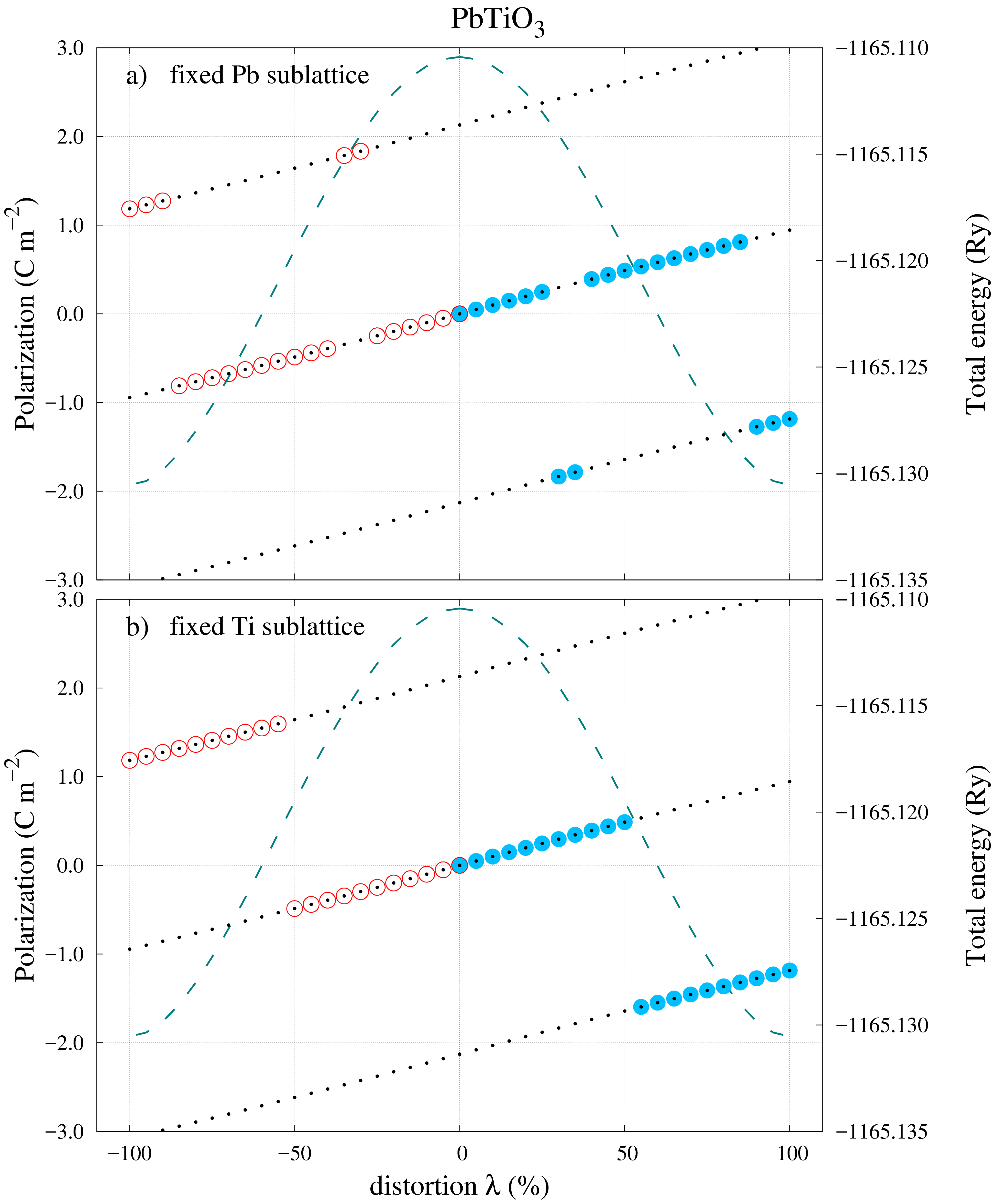}
    \caption{Polarization lattice for $PbTiO_3$ (black dots given by Eq. \eqref{eq:pstate}) with distortion from the CS to the ``DOWN" FS (red empty circles) and from the CS to the ``UP" FS (solid blue circles) with respect to the $Pb$ sublattice a) and with respect to the $Ti$ sublattice b). The distortion parameter $\lambda$ was modified in $5\%$ steps, with the negative values for the ``DOWN" intended for a better clarity of the figure. The green dashed line is the total energy as a function of distortion showing a continuous variation for each distortion path.}
    \label{fig:pto}
\end{figure}

For the first distortion path in Figure \ref{fig:pto}a) more polarization jumps appear not only towards the end of the distortion interval. 
This only reflects the ``random" nature of these polarization jumps that are driven by the numerical implementation of the BP polarization theory. 
In order to verify that these jumps are not related to any numerical errors, the total energy is plotted (dashed green line) as a function of the system distortion. 
If the irregularities in the dependence of polarization on the system distortion had a different source, a corresponding discontinuity in the energy dependence should be observed. 

This is clearly not the case in any of the studied examples. 
Apart from the appearance of irregular polarization jumps, the important FS and CS polarization are located on different branches similar to the case in Figure \ref{fig:bto}b): $P_{CS}=\SI{0}{\coulomb\per\metre\squared}$, $P_{FS}=\SI{-1.1852102}{\coulomb\per\metre\squared}$ and $P_{q}=\SI{2.13}{\coulomb\per\metre\squared}$. 
The spontaneous polarization can be obtained in a similar fashion, by adding one polarization quanta (for the ``UP" direction): $P_s=P_{FS}+P_q-P_{CS}=\SI{0.944}{\coulomb\per\metre\squared}$. 
It should be noted that the value of the obtained spontaneous polarization in this case is not ``much smaller" than the polarization quanta and this can add more confusion when analysing calculated data. 
For the second distortion path in Figure \ref{fig:pto}b), the irregular jumps from the previous case have disappeared but the polarization ambiguity is solved just like in the previous case. 
In fact, if it was not for the one instance in Figure \ref{fig:bto}a), one could be inclined to conclude at this point that, by always adding one polarization quanta to $P_{FS}$ for the ``UP" direction, Eq. \eqref{eq:pss} would lead to an unambiguous result. 
Unfortunately this is not true, as the next calculated example will show. 
%--------------------------------------------------------------------------------------------------------
\subsection*{$KNbO_3$}
\begin{figure}[h]
    \centering
    \includegraphics[width=\columnwidth]{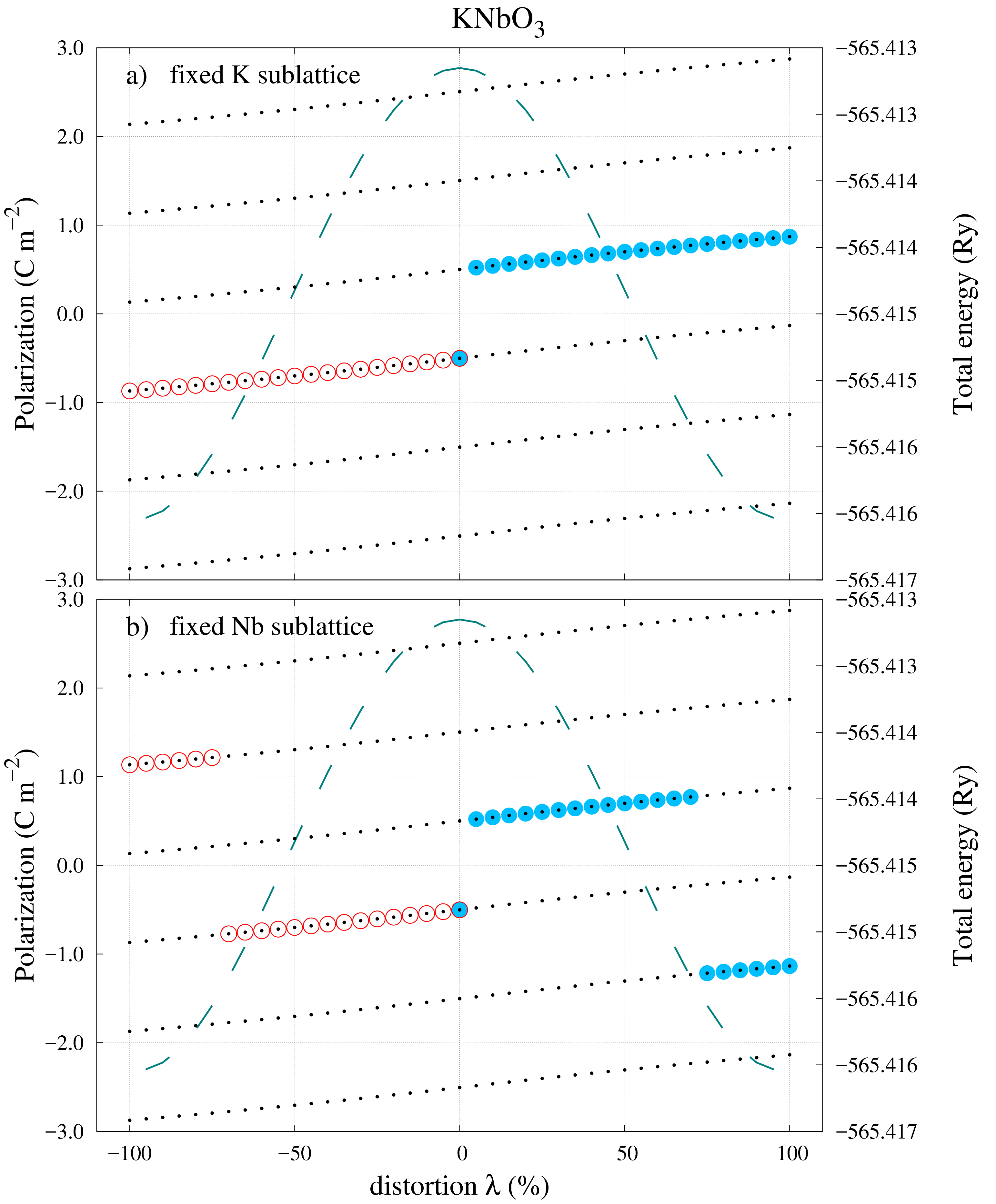}
    \caption{Polarization lattice for $KNbO_3$ (black dots given by Eq. \eqref{eq:pstate}) with distortion from the CS to the ``DOWN" FS (red empty circles) and from the CS to the ``UP" FS (solid blue circles) with respect to the $K$ sublattice a) and with respect to the $Nb$ sublattice b). The distortion parameter $\lambda$ was modified in $5\%$ steps, with the negative values for the ``DOWN" intended for a better clarity of the figure. The green dashed line is the total energy as a function of distortion showing a continuous variation for each distortion path.}
    \label{fig:kno}
\end{figure}

Figure \ref{fig:kno} shows the same type of polarization calculations as the previous two cases, however the results are entirely different, starting with the CS polarization value. 
Contrary to the results in Figures \ref{fig:bto} and \ref{fig:pto} the CS polarization in Figure \ref{fig:kno} does not vanish for the centrosymmetric state. 
This result appears to be in contradiction with the expectation that for a centrosymmetric state of a system the polarization must vanish, yet this is only true if the polarization was single valued! \cite{rabe2007,spaldin2012}. 
Once again it is shown that using the current approach it is impossible to successfully apply the same strategy to different materials. 
Another difference in the results obtained for KNO is the polarization jump that appears in the near vicinity of the CS. 
This only appears for one polarization direction but not the other. 
This is a fortuitous result because it reinforces the fact that in order for Eq. \eqref{eq:pss} to be applied, both the CS and the FS polarizations must belong to the same branch.
For this case the CS and FS values for the ``UP" direction in Figure \ref{fig:kno}a) were: $P_{CS}=\SI{-0.501}{\coulomb\per\metre\squared}$, $P_{FS}=\SI{0.870}{\coulomb\per\metre\squared}$ and $P_{q}=\SI{1.002}{\coulomb\per\metre\squared}$. 
By correcting either value, the spontaneous polarization can be obtained to be $P_s=\SI{0.369}{\coulomb\per\metre\squared}$. 
%--------------------------------------------------------------------------------------------------------
\section{\label{sec:optimization}Optimization strategy for Berry phase polarization calculations}
The examples shown in Figures \ref{fig:bto}-\ref{fig:kno} have been presented using a large number of calculations in order to help visualise polarization branches. 
In a regular scenario, some cases can be clarified using a smaller number of calculations. 
On average, depending on the choice of the distortion parameter value $\lambda$, four calculations should be enough to correctly identify the polarization branches and obtain the spontaneous polarization for the first three cases presented in Section \ref{sec:calc}. 
Yet the only common feature that remains is that each material requires an individual treatment to obtain the spontaneous polarization. 
This process is far more tedious for larger system sizes. 
In order to illustrate our optimized method we show the case of PZT. 
\begin{figure}[h]
    \centering
    \includegraphics[width=\columnwidth]{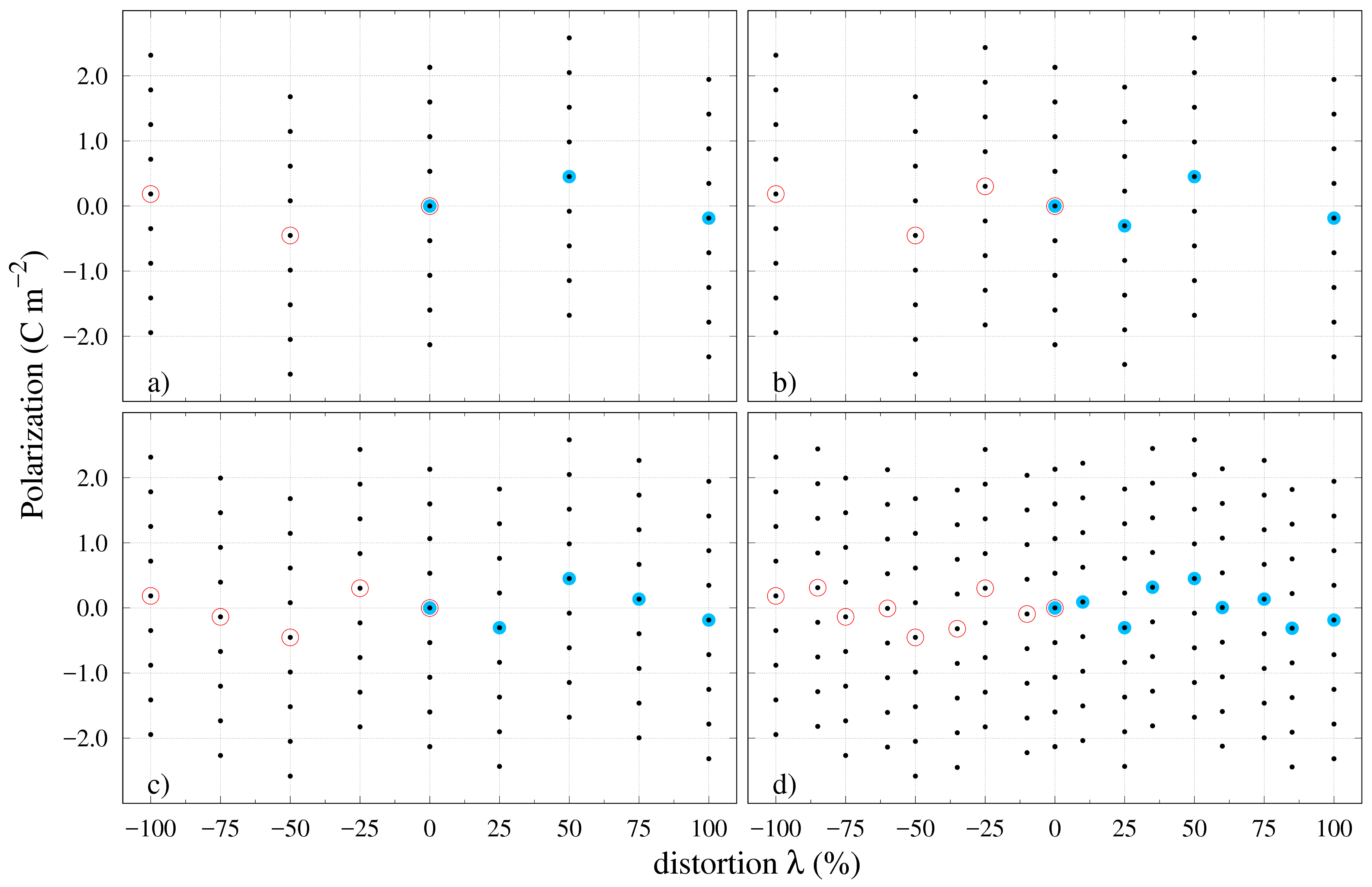}
    \caption{Polarization lattice for $Pb(Zr_{0.25}Ti_{0.75})O_3$ super-cell (black dots) with the calculated values in filled blue circles for the ``UP" polarization direction and empty red circles for ``DOWN": a) one intermediary point at $50\%$ distortion; b) two intermediary points: $25\%$ and $50\%$ distortion; c) three intermediary points: $25\%$, $50\%$ and $75\%$ distortion; d) seven intermediary points: $10\%$, $25\%$, $35\%$, $50\%$, $60\%$, $75\%$ and $85\%$ distortion.}
    \label{fig:pzt}
\end{figure}

The polarization calculations for PZT are shown in Figure \ref{fig:pzt}a)-d) for an ever increasing number of intermediary system distortion states. 
Due to the site and the large number of atoms in the super-cell it becomes difficult to fix a sublattice and perform the distortion with respect to it. 
As an alternative, in this case, all atoms in the unit cell have been moved as described in Eq. \eqref{eq:distort} from the CS to the corresponding FS. 
The calculated polarization values for the CS and FS are: $P_{CS}=\SI{0.0}{\coulomb\per\metre\squared}$, $P_{FS}=\SI{-0.186}{\coulomb\per\metre\squared}$ and the polarization quanta $P_{q}=\SI{0.532}{\coulomb\per\metre\squared}$. 
This seems to be an ambiguous situation and generally for an unknown material it is difficult to know for sure just from the difference between the FS and CS polarization values. 
For this reason, an extra point has been introduced in Figure \ref{fig:pzt}a) corresponding to a $50\%$ system distortion. 
The purpose for adding more intermediary states is to obtain a clear image of the polarization branches in order to identify what correction is needed for the values of interest. 
In this case, the extra point does not help to clarify the situation completely since no polarization branches are revealed and we make another choice for the next system distortion. 
This time, a $25\%$ distortion is added to the plot in Figure \ref{fig:pzt}b) and the result does not bring the conclusion any closer. 
Moving on, a $75\%$ distortion is added next in Figure \ref{fig:pzt}c) with the same result. 
It is only when smaller distortion steps are added in Figure \ref{fig:pzt}d) that a real image of the PZT polarization branches emerges and a correction can be made by adding two polarization quanta to the FS value in order to bring it on the same branch as the CS case. 

Figure \ref{fig:pzt}d) provides the best clue to introducing an optimization procedure. 
The image that can be formed from all the results presented in this study is that the polarization values belonging to the same branch will have a monotonous increase in \emph{absolute value} starting from the reference state towards the ferroelectric one. 
This result has been discussed at length by Resta et. al \cite{resta1993a,rabe2007} and it is known as the linear response theory, which states that the variation of polarization on the internal distortion should be linear. 
This has been shown to be true for KNO \cite{resta1993a} and it can be observed to hold for the rest of the examples in this study, however it does not apply to all materials in general. 
Similar polarization studies on $BiFeO_3$, reveal a deviation from the linear dependence beyond $\approx20\%$ distortion \cite{neaton2005,rabe2007,spaldin2012}. 
Nevertheless, in most cases one could safely use the following approximation for the polarization dependence on the unit-cell distortion parameter:
\begin{equation}
P_{lin}(\lambda)=\Lambda \, \lambda+P_{CS},
\label{eq:plin}
\end{equation}
where, $P_{CS}$ is the \emph{calculated} polarization of the reference state of the studied system, which in the examples of this study was the centrosymmetric state and $\Lambda$ is the slope of the linear response. 
In order to be able to use this approximation the slope parameter $\Lambda$ must be computed through a ``finite difference" route:
\begin{equation}
\Lambda=\frac{P_{d\lambda}-P_{CS}}{d\lambda},
\label{eq:lambda}
\end{equation}
where $P_{d\lambda}$ is the \emph{calculated} polarization for an ``infinitesimal" system distortion $d\lambda$ in the vicinity of the reference CS. 
However, from the definition of polarization in the BP theory, both polarization values in Eq. \eqref{eq:lambda} are determined up to an integer value of polarization quanta. 
\begin{figure}[h]
    \centering
    \includegraphics[width=\columnwidth]{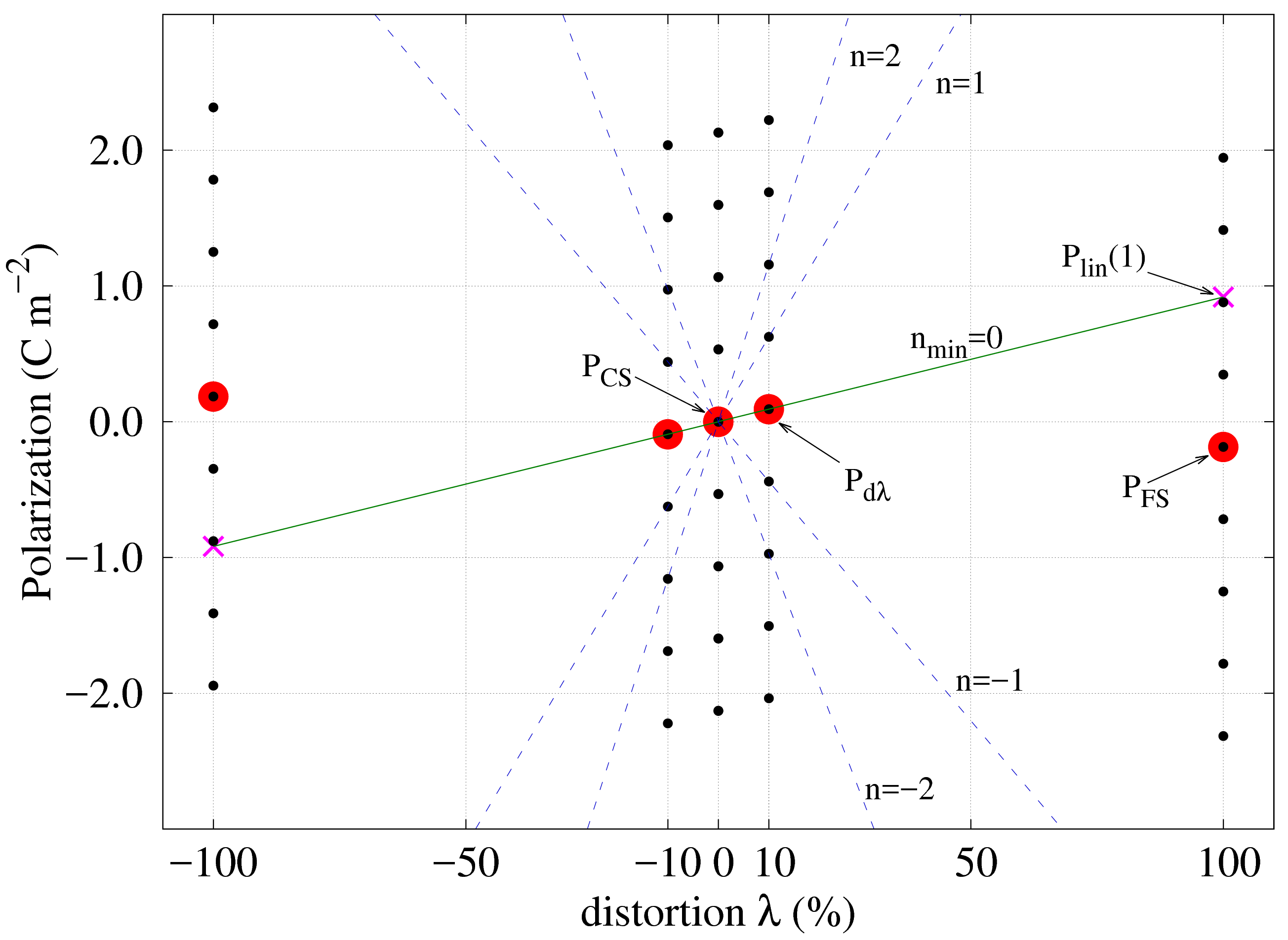}
    \caption{Demonstrating the optimization procedure for finding the correct polarization branch for PZT in the ``UP" direction. 
        Red circles represent calculated points while the black points are the corresponding polarization lattice. 
        The green line is the linear approximation defined in Eq. \eqref{eq:plin}. 
        The blue dashed lines exemplify the linear approximation for a slope parameter different than the minimal value obtained for $n_{min}$. 
        The magenta coloured crosses are the interpolated values for the two FS.}
    \label{fig:fit}
\end{figure}
Therefore the difference is still multivalued and, by extension, the slope parameter has a similar behaviour. 
The problem can be solved if the entire family of values for $P_{d\lambda}$ is introduced in Eq. \eqref{eq:lambda}, while the polarization for the reference state is kept at the calculated value $P_{CS}$ and its respective branch:
\begin{equation}
\Lambda(n)=\frac{P_{d\lambda}+n \, P_q - P_{CS}}{d\lambda}, \, n \in \mathbb{Z}.
\label{eq:lambdaK}
\end{equation}

It is now possible to find the integer number $n_{min}$ such that $\left|\Lambda(n_{min})\right|$ in Eq. \eqref{eq:lambdaK} is minimum. 
This last operation essentially ensures that the calculated polarization for the ``infinitesimal" distortion $d\lambda$ is brought to the same branch as the polarization for the reference CS. 
This can be easily verified by calculating several values for $\Lambda(n)$ and then using Eq. \eqref{eq:plin} to plot the corresponding dependencies. 
Figure \ref{fig:fit} shows the linear polarization dependence for different values of the branch index $n$ corresponding to the calculated $P_{d\lambda}$ polarization. 
The continuous green line is obtained for the minimum $\abs{\Lambda(n_{min})}$, where $n_{min}=0$. 
Using $n_{min}$, Eq. \eqref{eq:lambdaK} returns the slope of the linear dependence of polarization independent of branch! 
With this information, Eq. \eqref{eq:plin} can be used to extrapolate the polarization value at $100\%$ distortion in the FS and subsequently obtain the integer number $k$ of quanta to correct the branch ambiguity:
\begin{equation}
k=\left[ \frac{P_{lin}(\lambda=100\%)-P_{FS}}{P_q} \right], \, k \in \mathbb{Z},
\label{eq:kapp}
\end{equation}
where, $P_{FS}$ is the calculated polarization. 
Finally, the spontaneous polarization can be obtained using Eq. \eqref{eq:pss} and the corrected $P_{FS}$ value:
\begin{equation}
P_s=(P_{FS}+k\,P_q)-P_{CS}.
\label{eq:pspontaneous}
\end{equation}

All the steps of the proposed optimization procedure have been summarized in a simple work flow in Figure \ref{fig:optimwfl}. 
For materials such as the ones presented in this study, where the polarization direction is parallel to one of the principal axes, the total number of calculations is reduced to only three system states (steps 1, 3 and 7). 
The rest of the steps can be performed with simple arithmetic operations that can be easily included in a high throughput calculation automated script. 
For a more general case, the ferroelectric polarization can take any direction in the crystal and the operations proposed by this strategy must be repeated for each of the three cartesian axes. 
%--------------------------------------------------------------------------------------------------------
\section{\label{sec:res}Results}
\begin{figure}[h]
    \centering
    \includegraphics[width=\columnwidth]{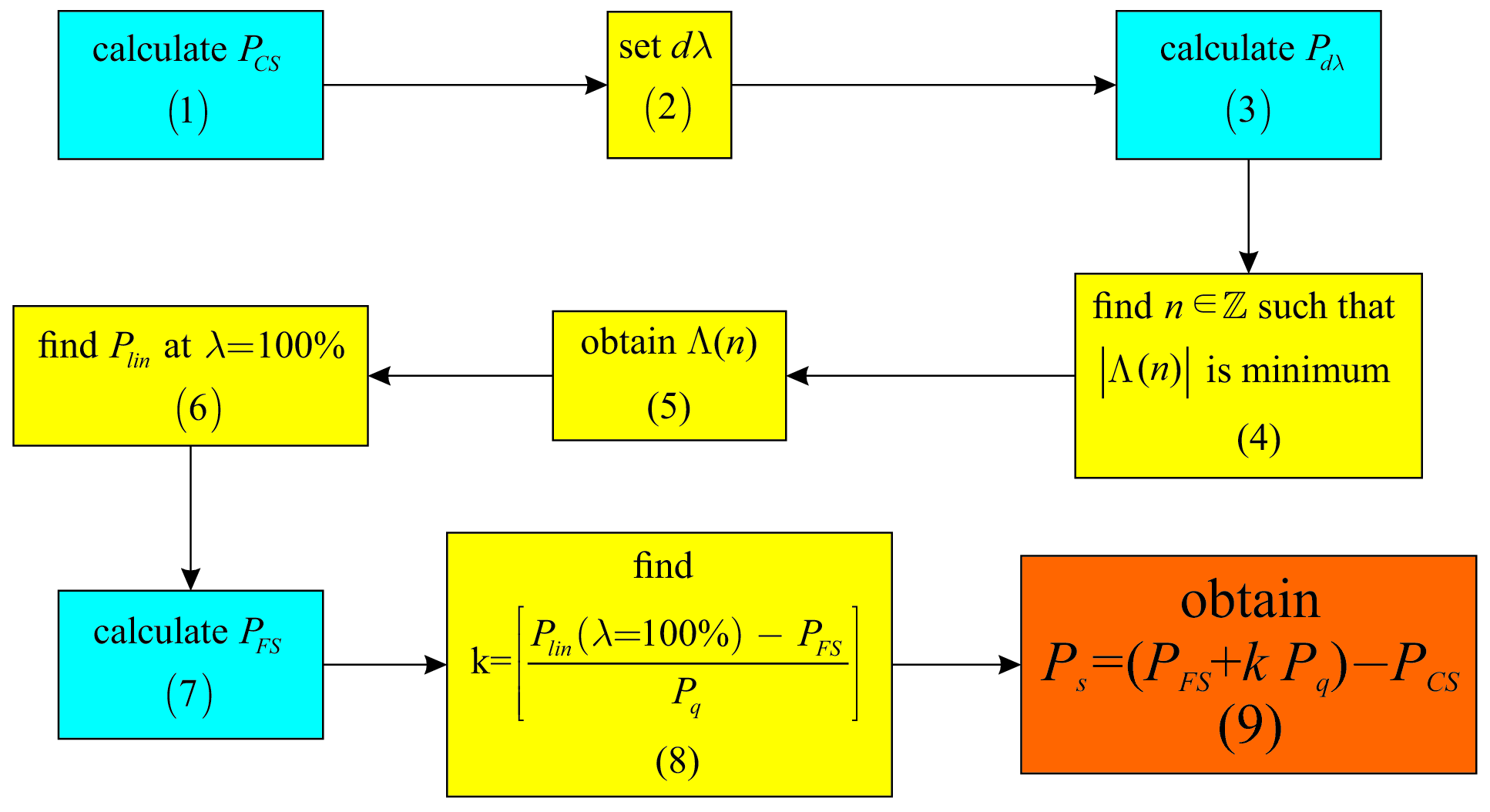}
    \caption{Optimized work-flow for Berry phase calculations. The blue rectangles are polarization calculation runs and in yellow and orange are regular arithmetic operations.}
    \label{fig:optimwfl}
\end{figure}
The most important result obtained in this study is the optimization strategy summarized in Figure \ref{fig:optimwfl}. 
By comparison with the original method for Berry phase calculations in Figure \ref{fig:berrywfl} the proposed approach does not contain any interrogations steps and algorithm branches. 
Also, the method is a series of nine consecutive steps and only three calculation runs. 
The original approach however, may need considerably more calculations if multiple polarization jumps occur for the chosen distortion path of the system. 
Analysing Eq. \eqref{eq:lambda} more closely, one can recognise a similitude with the method for calculating Born effective charges \cite{spaldin2012,rabe2007,resta1993a}. 
By definition, Born effective charges represent the change in polarization divided by how much an ion is displaced \cite{spaldin2012,rabe2007,resta1993a} which translates to the first derivative of polarization with respect to atomic displacement while keeping the rest of the ions and external macroscopic fields fixed \cite{resta1993a}. 
In a practical calculation, this definition is transformed into a finite difference formula. 
Considering one of the materials studied in this paper, these charges are calculated by starting with the CS and then displacing atom $X$ by a small amount (typically around $\SI{0.005}{\angstrom}$) along the polarization direction, the Berry phase method is used to calculate the polarization. 
The Born effective charge associated to atom $X$ for a displacement along polarization direction is obtained as:
\begin{equation}
Z^{*}_{X}=\frac{\Omega}{e} \, \frac{P(dr)-P(0)}{dr},
\label{eq:born}
\end{equation}
where $dr$ is the infinitesimal displacement of atom $X$, $\Omega$ is the volume of the unit-cell and $e$ is the electron charge. 
After computing all the Born effective charges for the atoms in the unit-cell then the following linear approximation for the spontaneous polarization can be written:
\begin{equation}
P=\frac{e}{\Omega} \, \sum\limits_{i}Z^{*}_{i}\,z_i,
\label{eq:Pborn}
\end{equation}
where the sum is over all the atoms in the unit cell and $z_i$ is the displacement of each atom in the FS (for this case, the displacements are only along the $z$ axis). 
It can be observed that the polarization approximation in Eq. \eqref{eq:plin} and Eq. \eqref{eq:lambda} follows a similar approach, but with the infinitesimal simultaneous displacement of all atoms. 

\begin{table*}[t]
    \renewcommand{\arraystretch}{1.5}
    \centering
    \begin{ruledtabular}
        \begin{tabular}{@{}cccccccccccc@{}}
            \multicolumn{3}{l}{} & $P_{CS}$ (\SI{}{\coulomb\per\metre\squared}) & $d\lambda$ & $P_{d\lambda}$ (\SI{}{\coulomb\per\metre\squared}) & $n$ & $\Lambda$ (\SI{}{\coulomb\per\metre\squared}) & $P_{lin}(1)$ (\SI{}{\coulomb\per\metre\squared}) & $P_{FS}$ (\SI{}{\coulomb\per\metre\squared}) & $k$ & $P_s$ (\SI{}{\coulomb\per\metre\squared}) \\
            \parbox[t]{1mm}{\multirow{7}{*}{\rotatebox[origin=c]{90}{DOWN}}} & \multirow{2}{*}{BTO} & Ba  & $0.000$ & $5\%$  & $-0.019$ & $0$ & $-0.0038$ & $-0.380$ & $-0.350$ & $0$ & \multirow{2}{*}{-0.350} \\
            &                                          & Ti  & $0.000$ & $5\%$ & $-0.019$ & $0$ & $-0.0038$ & $-0.380$ & $1.685$ & $-1$ &  \\ \cline{3-12}
            & \multirow{2}{*}{PTO}                     & Pb  & $0.000$ & $5\%$ & $-0.050$ & $0$ & $-0.010$ & $-1.000$ & $1.185$ & $-1$ & \multirow{2}{*}{$-0.945$} \\
            &                                          & Ti  & $0.000$ & $5\%$ & $-0.050$ & $0$ & $-0.010$ & $-1.000$ & $1.185$ & $-1$ &  \\ \cline{3-12}
            & \multicolumn{1}{c}{\multirow{2}{*}{KNO}} & K   & $-0.501$ & $5\%$ & $-0.522$ & $0$ & $-0.0042$ & $-0.921$ & $-0.870$ & $0$ & \multirow{2}{*}{$-0.369$} \\
            & \multicolumn{1}{c}{}                     & Nb  & $-0.501$ & $5\%$ & $-0.522$ & $0$ & $-0.0042$ & $-0.921$ & $1.135$ & $-2$ &  \\ \cline{3-12}
            & PZT                                      & None & $0.000$ & $10\%$ & $-0.093$ & $0$ & $-0.0093$ & $-0.930$ & $0.186$ & $-2$ & $-0.878$  \\ \cline{1-12}
            \parbox[t]{1mm}{\multirow{7}{*}{\rotatebox[origin=c]{90}{UP}}} & \multirow{2}{*}{BTO} & Ba & $0.000$ & $5\%$ & $0.019$ & $0$ & $0.0038$ & $0.380$ & $0.350$ & $0$ & \multirow{2}{*}{$0.350$} \\
            &                                          & Ti  & $0.000$ & $5\%$ & $0.019$ & $0$ & $0.0038$ & $0.380$ & $-1.685$ & $1$ &  \\ \cline{3-12}
            & \multirow{2}{*}{PTO}                     & Pb  & $0.000$ & $5\%$ & $0.050$ & $0$ & $0.010$ & $1.000$ & $-1.185$ & $1$ & \multirow{2}{*}{$0.945$} \\
            &                                          & Ti  & $0.000$ & $5\%$ & $0.050$ & $0$ & $0.010$ & $1.000$ & $-1.185$ & $1$ &  \\ \cline{3-12}
            & \multirow{2}{*}{KNO}                     & K   & $-0.501$ & $5\%$ & $0.522$ & $-1$ & $0.0042$ & $-0.081$ & $0.870$ & $-1$ & \multirow{2}{*}{$0.369$} \\
            &                                          & Nb  & $-0.501$ & $5\%$ & $0.522$ & $-1$ & $0.0042$ & $-0.081$ & $-1.134$ & $1$ &  \\ \cline{3-12}
            & PZT                                      & None & $0.000$ & $10\%$ & $0.093$ & $0$ & $0.0093$ & $0.93$ & $-0.186$ & $2$ & $0.878$ \\
        \end{tabular}
    \end{ruledtabular}
    \caption{Spontaneous polarization obtained using the proposed optimization strategy for all the studied materials shown in Figures \ref{fig:bto}-\ref{fig:pzt}.}
    \label{tab:results}
\end{table*}
Using the steps summarized in Figure \ref{fig:optimwfl} one can obtain the spontaneous polarization for all the cases presented in this study in Figures \ref{fig:bto}-\ref{fig:pzt} using only three of the calculated points for each distortion path. 
The results can be analysed in Table \ref{tab:results} where the values obtained for each step are shown separated between the two polarization directions (formally denoted as ``UP" and ``DOWN"). 
The third column of Table \ref{tab:results} indicates the sublattice with respect to which the systems were distorted. 
As described in Section \ref{sec:compdet}, for the PZT case all atoms were moved from their corresponding centrosymmetric positions in the super-cell toward the FS positions following the same rule in Equation \eqref{eq:distort}. 

The application of the proposed optimization strategy is straightforward yet there are some aspects that should be pointed out. 
First of all, the optimization strategy is based on the linear response theory of polarization as a function of the system distortion as discussed by Resta \emph{et. al} \cite{resta1993a}. 
This was shown to be true for the KNO case \cite{resta1993a} and it was shown in this work that similar results are obtained for PTO and BTO. 
However, in a study by Neaton \emph{et. al} on $BiFeO_3$ it is shown that the polarization dependence on the system distortion is only linear to about $20\%$ distortion. 
For situations where the materials have strong non-linear polarization dependence on the internal distortion, the proposed optimization strategy may need more calculated points in order to obtain a non-ambiguous result. 
Another important point to be considered, is one of the crucial conditions for the application of the BP theory: the system must be insulating in any state on the distortion path. 
This aspect is difficult to control,and such materials will require a more detailed investigation. 
%--------------------------------------------------------------------------------------------------------
\section{\label{sec:conc}Conclusions}
In summary, the results presented in the current study provide a detailed view of the numerical implementation of the modern theory of polarization from the point of view of actual calculations on various ferroelectric materials. 
Most of the studies have been purposely performed with an exaggerated fine distortion mesh in order to provide a better visual image of the multivalued aspect of polarization and the emergence of polarization branches. 
We illustrate this using $BaTiO_3$, $PbTiO_3$ and $KNbO_3$ as test materials. 
It has been shown that using the approach proposed in the usual Berry phase polarization calculations, each material must be treated individually which makes it difficult to integrate such a study in an automated work-flow.  
For this reason, a unified optimization procedure has been proposed that can provide a starting point for polarization investigation for ferroelectric materials. 
The procedure uses a minimal number of calculations in order to obtain the spontaneous polarization value, thus reducing the computational effort. 
We hope the present study will enhance the current efforts in the theoretical investigation of known ferroelectrics and accelerate the design of new ferroelectric materials. 
The current study and the proposed calculation strategy provides both a visual representation of the multivalued aspect of polarization in the modern theory and a practical approach for such calculations which should complement the theoretical descriptions found in the specialized literature. 
%--------------------------------------------------------------------------------------------------------
\section*{Acknowledgments}
The authors would like to acknowledge the useful discussions with Dr. Claude Ederer from ETH, Zurich that helped clarify the theoretical and practical aspects of the Berry phase polarization theory.

\noindent
LDF would like to acknowledge the financial support of the Romanian Ministry of Education Executive Unit for Funding High Education, Research, Development and Innovation (MEN-UEFISCDI) through the Young Research Team Grant PNII-RU-TE-2012-3-0320 (Contract No. 11).

\noindent
The authors would also like to acknowledge the financial support from the NIMP Core Program nr. PN16-480102.

%--------------------------------------------------------------------------------------------------------
%\section*{References}
%
\bibliographystyle{unsrt}
\bibliography{references}

\begin{thebibliography}{10}

\bibitem{scott1989}
J.~F. Scott and C.~A.~P. de~Araujo.
\newblock {\em Science}, 246:4936, 1989.

\bibitem{hidaka1992}
T.~Hidaka.
\newblock {\em Ferroelectrics}, 137(1):291--295, 1992.

\bibitem{glinchuk2009}
M.~D. Glinchuk, E.~V. Kirichenko, V.~A. Stephanovich, and B.~Y. Zaulychny.
\newblock {\em J. Appl. Phys.}, 105(10):104101, 2009.

\bibitem{kumari2015}
P.~Kumari, R.~Rai, S.~Sharma, M.~Shandilya, and A.~Tiwari.
\newblock {\em Adv. Mat. Lett.}, 6(6):453--484, 2015.

\bibitem{rong2016}
Y.~Rong, M.~Li, J.~Chen, M.~Zhou, K.~Lin, L.~Hu, W.~Yuan, W.~Duan, J.~Deng, and
  X.~Xing.
\newblock {\em Phys. Chem. Chem. Phys.}, 18(8):6247--6251, 2016.

\bibitem{scott2007}
J.~F. Scott.
\newblock {\em Science}, 315(5814):954--959, 2007.

\bibitem{choi2009}
T.~Choi, S.~Lee, Y.~J. Choi, V.~Kiryukhin, and S.-W. Cheong.
\newblock {\em Science}, 324(5923):63--66, 2009.

\bibitem{garcia2009}
V.~Garcia, S.~Fusil, K.~Bouzehouane, S.~Enouz-Vedrenne, N.~D. Mathur,
  A.~Barthélémy, and M.~Bibes.
\newblock {\em Nature}, 460(7251):81--84, 2009.

\bibitem{huang2010}
H.~Huang.
\newblock {\em Nat. Photon.}, 4(3):134--135, 2010.

\bibitem{yuan2011}
Y.~Yuan, T.~J. Reece, P.~Sharma, S.~Poddar, S.~Ducharme, A.~Gruverman, Y.~Yang,
  and J.~Huang.
\newblock {\em Nat. Mater.}, 10(4):296--302, 2011.

\bibitem{liu2014}
F.~Liu, W.~Wang, L.~Wang, and G.~Yang.
\newblock {\em Appl. Phys. Lett.}, 104(10):103907, 2014.

\bibitem{chen2015}
B.~Chen, X.~Zheng, M.~Yang, Y.~Zhou, S.~Kundu, J.~Shi, K.~Zhu, and S.~Priya.
\newblock {\em Nano Energy}, 13:582--591, 2015.

\bibitem{kim2016}
W.~Y. Kim, H.-D. Kim, T.-T. Kim, H.-S. Park, K.~Lee, H.~J. Choi, S.~H. Lee,
  J.~Son, N.~Park, and B.~Min.
\newblock {\em Nat. Commun.}, 7:10429, 2016.

\bibitem{moure2015}
C.~Moure and O.~Peña.
\newblock {\em Prog. Solid State Chem.}, 43(4):123--148, 2015.

\bibitem{pintilie2007}
L.~Pintilie, I.~Vrejoiu, G.~Le~Rhun, and M.~Alexe.
\newblock {\em J. Appl. Phys.}, 101(6):064109, 2007.

\bibitem{kundu2015}
S.~Kundu, M.~Clavel, P.~Biswas, B.~Chen, H.-C. Song, P.~Kumar, N.~N. Halder,
  M.~K. Hudait, P.~Banerji, M.~Sanghadasa, and S.~Priya.
\newblock {\em Sci. Rep.}, 5:12415, 2015.

\bibitem{volonakis2016}
G.~Volonakis, M.~R. Filip, A.~A. Haghighirad, N.~Sakai, B.~Wenger, H.~J.
  Snaith, and F.~Giustino.
\newblock {\em J. Phys. Chem. Lett.}, 7(7):1254--1259, 2016.

\bibitem{filip2016}
M.~R. Filip and F.~Giustino.
\newblock {\em J. Phys. Chem. C}, 120(1):166--173, 2016.

\bibitem{filip2016a}
M.~R. Filip, S.~Hillman, A.~A. Haghighirad, H.~J. Snaith, and F.~Giustino.
\newblock {\em J. Phys. Chem. Lett.}, 7(13):2579--2585, 2016.

\bibitem{berry1984}
M.~V. Berry.
\newblock {\em Proceedings of the Royal Society of London. A. Mathematical and
  Physical Sciences}, 392(1802):45--57, 1984.

\bibitem{resta1992}
R.~Resta.
\newblock {\em Ferroelectrics}, 136(1):51--55, 1992.

\bibitem{kingsmith1993}
R.~King-Smith and D.~Vanderbilt.
\newblock {\em Phys. Rev. B}, 47(3):1651--1654, 1993.

\bibitem{resta1993a}
R.~Resta, M.~Posternak, and A.~Baldereschi.
\newblock {\em Phys. Rev. Lett.}, 70(7):1010--1013, 1993.

\bibitem{resta1993b}
R.~Resta.
\newblock {\em EPL}, 22(2):133, 1993.

\bibitem{resta1994}
R.~Resta.
\newblock {\em Rev. Mod. Phys.}, 66(3):899--915, 1994.

\bibitem{rabe2007}
K.~M. Rabe, C.~H. Ahn, and J.-M. Triscone.
\newblock Topics in applied physics, 105. Springer, Berlin, 2007.

\bibitem{neaton2005}
J.~Neaton, C.~Ederer, U.~Waghmare, N.~Spaldin, and K.~Rabe.
\newblock {\em Phys. Rev. B}, 71(1), 2005.

\bibitem{hohenberg1964}
P.~Hohenberg and W.~Kohn.
\newblock {\em Phys. Rev.}, 136(3B):B864--B871, 1964.

\bibitem{giannozzi2009}
P.~Giannozzi, S.~Baroni, N.~Bonini, M.~Calandra, R.~Car, C.~Cavazzoni,
  D.~Ceresoli, G.~L. Chiarotti, M.~Cococcioni, I.~Dabo, A.~Dal~Corso,
  S.~de~Gironcoli, S.~Fabris, G.~Fratesi, R.~Gebauer, U.~Gerstmann,
  C.~Gougoussis, A.~Kokalj, M.~Lazzeri, L.~Martin-Samos, N.~Marzari, F.~Mauri,
  R.~Mazzarello, S.~Paolini, A.~Pasquarello, L.~Paulatto, C.~Sbraccia,
  S.~Scandolo, G.~Sclauzero, A.~P. Seitsonen, A.~Smogunov, P.~Umari, and R.~M.
  Wentzcovitch.
\newblock {\em J. Phys.: Cond. Matt.}, 21(39):395502, 2009.

\bibitem{perdew2008}
J.~P. Perdew, A.~Ruzsinszky, G.~I. Csonka, O.~A. Vydrov, G.~E. Scuseria, L.~A.
  Constantin, X.~Zhou, and K.~Burke.
\newblock {\em Phys. Rev. Lett.}, 100(13):136406, 2008.

\bibitem{theos}
\url{http://theossrv1.epfl.ch/Main/Pseudopotentials}.

\bibitem{monkhorst1976}
H.~J. Monkhorst and J.~D. Pack.
\newblock {\em Phys. Rev. B}, 13(12):5188, 1976.

\bibitem{wang2010}
J.~J. Wang, F.~Y. Meng, X.~Q. Ma, M.~X. Xu, and L.~Q. Chen.
\newblock {\em J. Appl. Phys.}, 108(3):034107, 2010.

\bibitem{saghiszabo1998}
G.~{S\'{a}ghi-Szab\'{o}}, R.~E. Cohen, and H.~Krakauer.
\newblock {\em Phys. Rev. Lett.}, 80(19):4321, 1998.

\bibitem{dallolio1997}
S.~Dall'Olio, R.~Dovesi, and R.~Resta.
\newblock {\em Phys. Rev. B}, 56(16):10105--10114, 1997.

\bibitem{spaldin2012}
N.~A. Spaldin.
\newblock {\em J. Solid State Chem.}, 195:2--10, 2012.

\end{thebibliography}
\newpage
%--------------------------------------------------------------------------------------------------------
\end{document}